\documentclass[aps,twocolumn,groupedaddress,amsmath,amssymb]{revtex4-1}
\usepackage{graphicx}  % needed for figures
\usepackage{dcolumn}   % needed for some tables
\usepackage{bm}        % for math
\usepackage{verbatim}   % for math

\begin{document}

\title{Self-organised-criticality and punctuated equilibrium in bouncing balls}

\author{Kaushal Gianchandani}
\email{kaushal.g@niser.ac.in}
\affiliation{National Institute of Science Education and Research Bhubaneswar, P.O. Jatni, Khurda - 752050, Odisha, India}

\author{A. N. Sekar Iyengar}
 \email{ansekar.iyengar@saha.ac.in}
\affiliation{Saha Institute of Nuclear Physics, Sector - 1, Block - AF Bidhan nagar, Kolkata - 700064, West Bengal, India}

\author{Prasanta K. Panigrahi}
\email{pprasanta@iiserkol.ac.in}
\affiliation{ 
Indian Institute of Science Education and Research Kolkata, Mohanpur - 741246, West Bengal, India}

\date{\today}

\begin{abstract}
A nonlinearly coupled system of bouncing balls is shown to exhibit features like self-organised-criticality (SOC) and punctuated equilibrium (PE) in suitable parameter domains. The temporal evolution of the non-stationary amplitudes is analysed through local methods to unravel the transient periodic components and fluctuations giving rise to SOC and PE type behaviours. This simple dynamical system follows Gutenberg-Richter relation and also manifests the Devil's staircase, explicating the universality of these features in diverse complex systems.
\end{abstract}

\maketitle

\section{Introduction}
Complex systems manifest a range of dynamic behaviour, many of which are common to widely disparate physical systems. Self-organised-criticality (SOC) \cite{bak1988self}, fractal and multi-fractal behaviour \cite{manimaran2009multiresolution}, synchronisation \cite{strogatz2001nonlinear, arenas2008synchronization}, amplitude death \cite{prasad2005amplitude},  self-organised motion in granular matter \cite{ramaswamy} are a few such examples, which have been successfully modelled.  Biological systems are conjectured to exhibit  ``punctuated equilibrium" \cite{1977}, where evolution is marked by sudden onsets followed by long periods of stasis.  
This feature manifested in a SOC based model \cite{PhysRevLett.71.4083}, which was later extend to the dynamics of earthquake fault motion \cite{PhysRevE.52.3232}. 
\par
The common appearance of these universal features indicate the possibility of modelling them in smaller systems, as compared to extremely complex biological or other physical systems. In this approach the origin of such behaviours can be made transparent. 
\par
Here, we demonstrate the manifestation of both SOC and PE type behaviour in the dynamics of multiple coupled bouncing balls. These are known to reveal nonlinear behaviour;  earlier studies involving a single ball have indicated the richness of its dynamics \cite{PhysRevLett.65.393, PhysRevE.48.3988, abhinna}. The analysis of the non-stationary amplitudes, carried out through local methods, revealed transient periodic components and fluctuations. In case of Devil's staircase, long periods of stasis with small jumps, spread wide apart in time were observed for a single ball. However, the jumps became more frequent and the period of stasis lessened with the increase in the number of balls. 
\par
\begin{figure}
\centering
\includegraphics[width = 0.45 \textwidth]{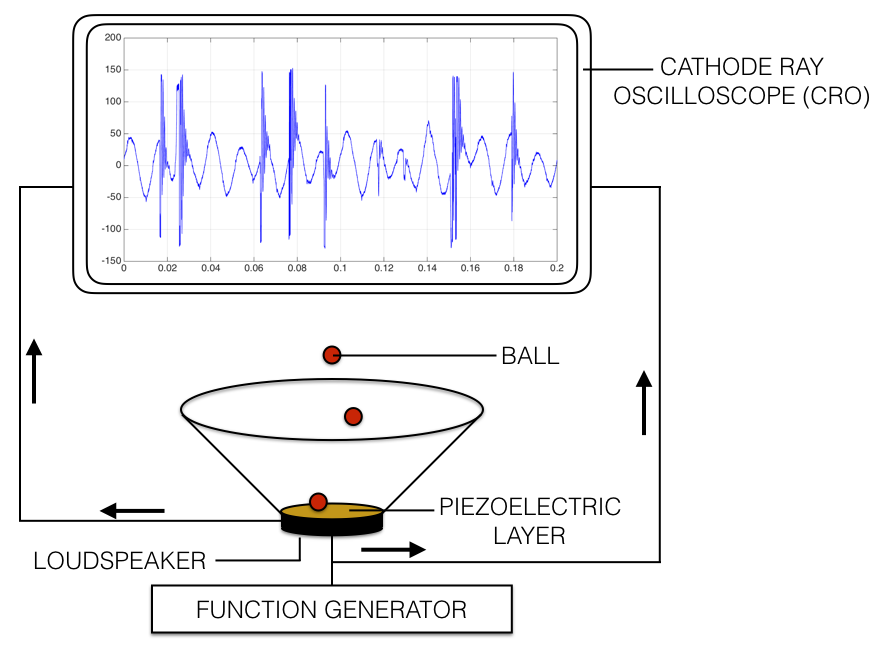}
\caption{\label{schematic_setup}Schematic for the experimental setup, with three bouncing balls, illustrating the output from the piezoelectric layer for input potential $12V$, $\omega = 80 Hz$.}
\end{figure}

The system under consideration is the classical bouncing ball experimental setup \cite{:/content/aapt/journal/ajp/67/3/10.1119/1.19229} with multiple balls bouncing on the piezoelectric layer. It consists of a speaker to which a sinusoidally vibrating input is provided using a function generator. The piezoelectric layer is fixed to the diaphragm of the speaker and its output is monitored through a cathode ray oscilloscope (CRO). In our experiment, the number of balls is increased in every iteration by one unit up to a maximum of five balls bouncing on the platform simultaneously. The balls are coupled to each other nonlinearly by a common base, through a variety of frequency couplings. The details of the masses and the radii of the balls are given in TABLE \ref{ball_m_r} and a schematic diagram of the setup is provided in FIG. \ref{schematic_setup}.

\begin{table}
\caption{\label{ball_m_r}Masses and radii of the bouncing balls}
\begin{ruledtabular}
\begin{tabular}{ccc}
Ball no.&Mass (in gms)&Radius (in mm)\\
\hline
1&1.0467&3.14\\
2&1.0472&3.12\\
3&1.0475&3.14\\
4&1.0455&3.13\\
5&1.0452&3.12\\
\end{tabular}
\end{ruledtabular}
\end{table}

We varied the forcing potential ($V$), the frequency of input ($\omega$) and number of balls ($n$) bouncing on the piezoelectric platform and observed the time series through the CRO, as depicted in FIG. \ref{schematic_setup}. Sudden behaviour changes in motion, amounting to delta function type response, are clearly evident in the time series.  It is observed that the time series exhibits periodic behaviour along with the presence of transient ruptures, decaying relatively quickly. 
Keeping in mind this non-stationary behaviour, we make use of the Morlet wavelet having a Gaussian window and sinusoidal sampling function \cite{daubechies1992ten}  to isolate the periodicity and transience \cite{abhinna}. The scalogram of the continuous wavelet transform (CWT) coefficients clearly depicts both these behaviours.
\begin{figure}
\centering
\includegraphics[width = 0.45 \textwidth]{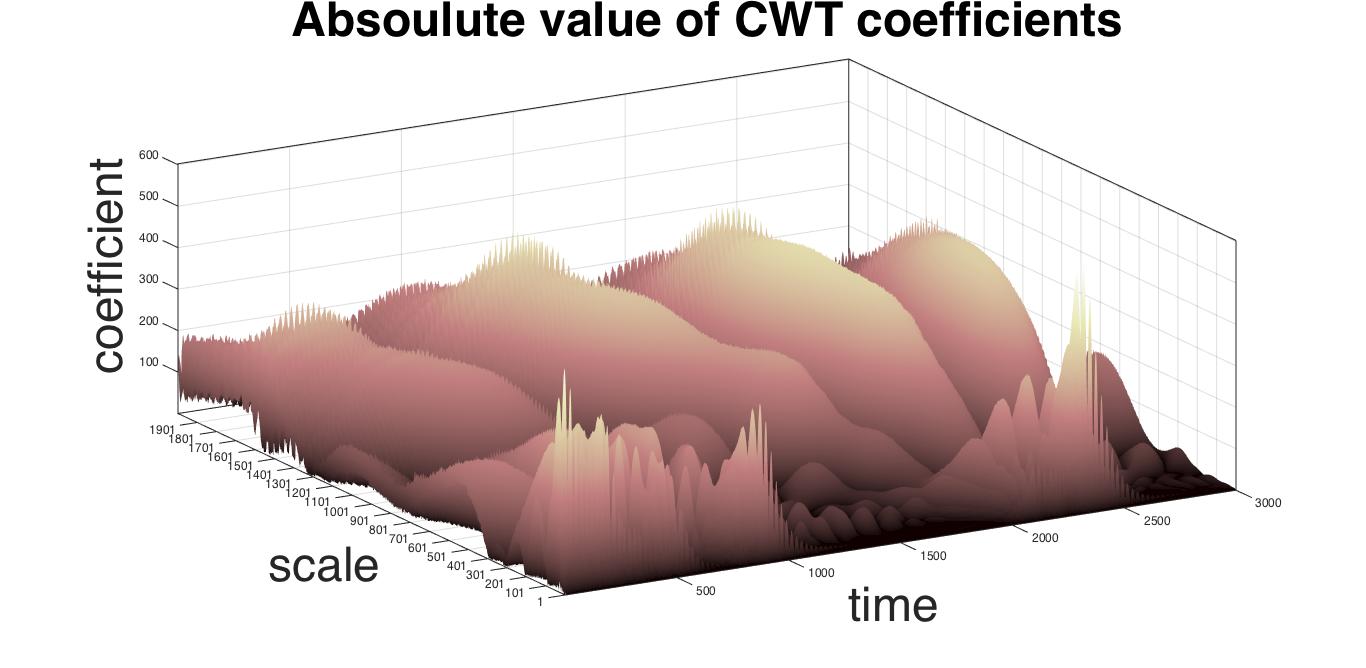}
\caption{\label{cwt}CWT coefficients revealing periodic and transient behaviour for a typical configuration of the system.}
\end{figure}

The over completeness of Morlet wavelets makes the quantitative analysis of the high frequency and the low frequency components difficult. It is worth mentioning that discrete wavelets can be used for extracting fluctuations \cite{manimaran2005wavelet}. Keeping in mind the fact that discrete wavelets are asymmetric and have their own fractal characteristics, we take the course to Hilbert-Huang transform (HHT) \cite{Huang08031998} for our analysis. HHT achieves the local separation of high and low frequency modulations by identifying the constituent  modes through Empirical Mode decomposition (EMD). Subsequently the time frequency localisation is achieved through a Hilbert transform. It has found effective application in analysing earthquake motion recordings \cite{zhang}. Below, we briefly outline the features of EMD before proceeding to our analysis.

\section{Empirical Mode Decomposition}
As is evident from physical considerations, a typical signal is a superposition of fast oscillations over slower ones and can be represented in terms of amplitude and frequency modulated components:
\begin{equation}
z(t) = \sum_{i = 1}^{N}c_{i}(t)\phi_{i}(t).
\end{equation}
Here, $c_{i}$s and $\phi_{i}$s are the amplitude and frequency modulated parts of the signal respectively. 
\par
The evolution of a signal can be better understood by studying its behaviour between two maxima, occurring at time $t_{1}$ and $t_{2}$. One can identify the localised higher-frequency components, lying inside the oscillation between $t_{1}$ and $t_{2}$. It is clear that the signal passes through these maxima and at least one minimum, which lies in the time interval $dt$: $\{dt | t_{1}<t<t_{2}\}$. 
\par
The signal can now be perceived as constituting of two parts, low frequency part $l(t)$ and the high-frequency part $h(t)$. This formalism can be implemented for all the oscillations which make up the signal, enabling one to iteratively isolate all the components of the signal. Evidently EMD is designed to consider oscillations in a signal at the local level in time. 
\par
The algorithm \cite{rilling2003empirical} for implementing EMD is described as follows:
\begin{enumerate}
\item Identify all extrema of $z(t)$.
\item Interpolate between the maxima and acquire an envelope $e_{max}(t)$. Similarly obtain the envelope $e_{min}(t)$ by interpolation between the minima.
\item Compute the mean $l(t) = (e_{max}(t) + e_{min}(t))/2$.
\item Extract the high-frequency detail. $h(t) = z(t) - l(t)$.
\item Iterate on the residual $l(t)$.
\end{enumerate}
While implementing it in practice, the above process has to be modified with a break clause, which is implemented when the average over the residue is negligible. The details obtained are referred to as Intrinsic Mode Functions (IMF). By construction, the number of extrema reduces after every iteration, thus ensuring the decomposition completes in a finite number of iterations. 

\begin{figure}
\centering
\includegraphics[width = 0.45 \textwidth]{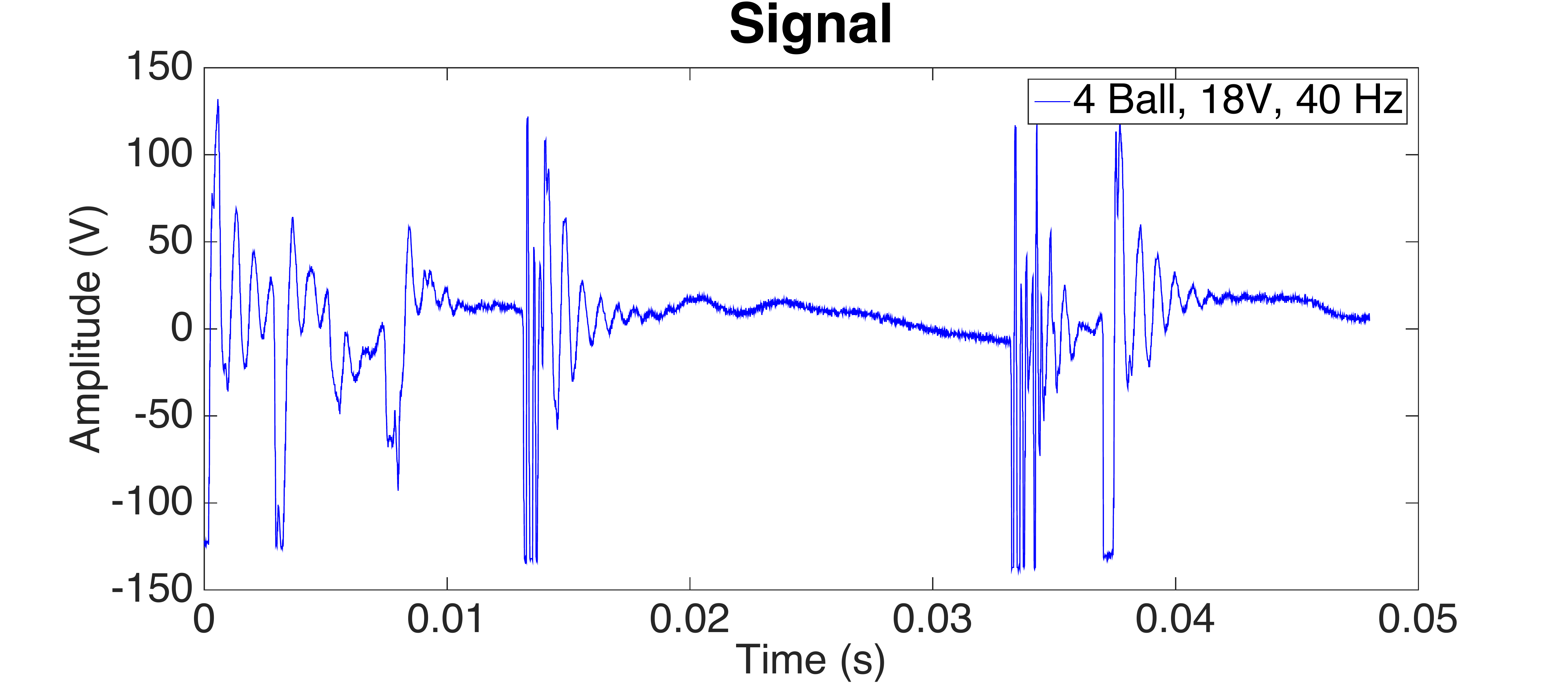}
\caption{\label{signal_emd}Signal for a typical configuration of the system, Butterworth filter was used for the purpose of de-noising the signal.}
\end{figure}

\begin{figure}
\centering
\includegraphics[width = 0.45 \textwidth]{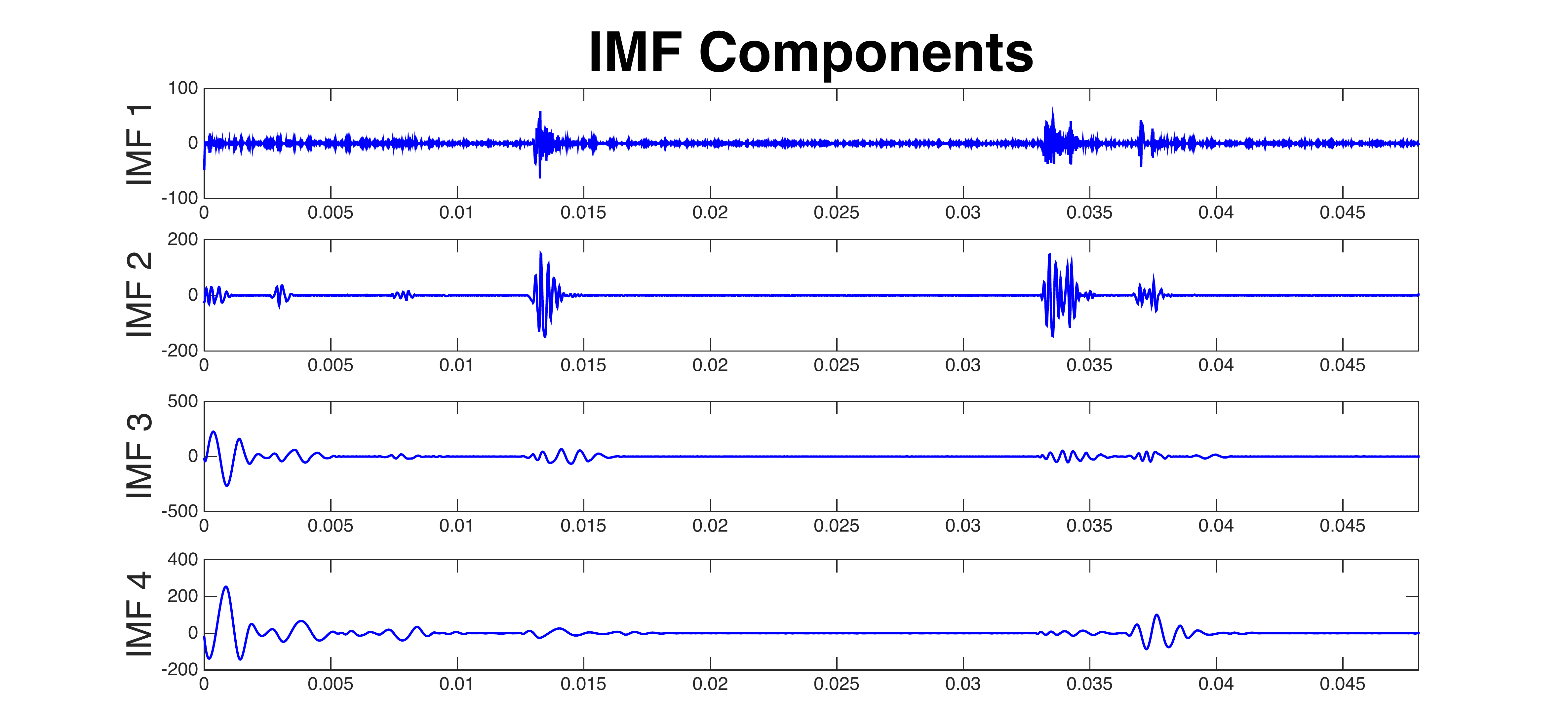}
\includegraphics[width = 0.45 \textwidth]{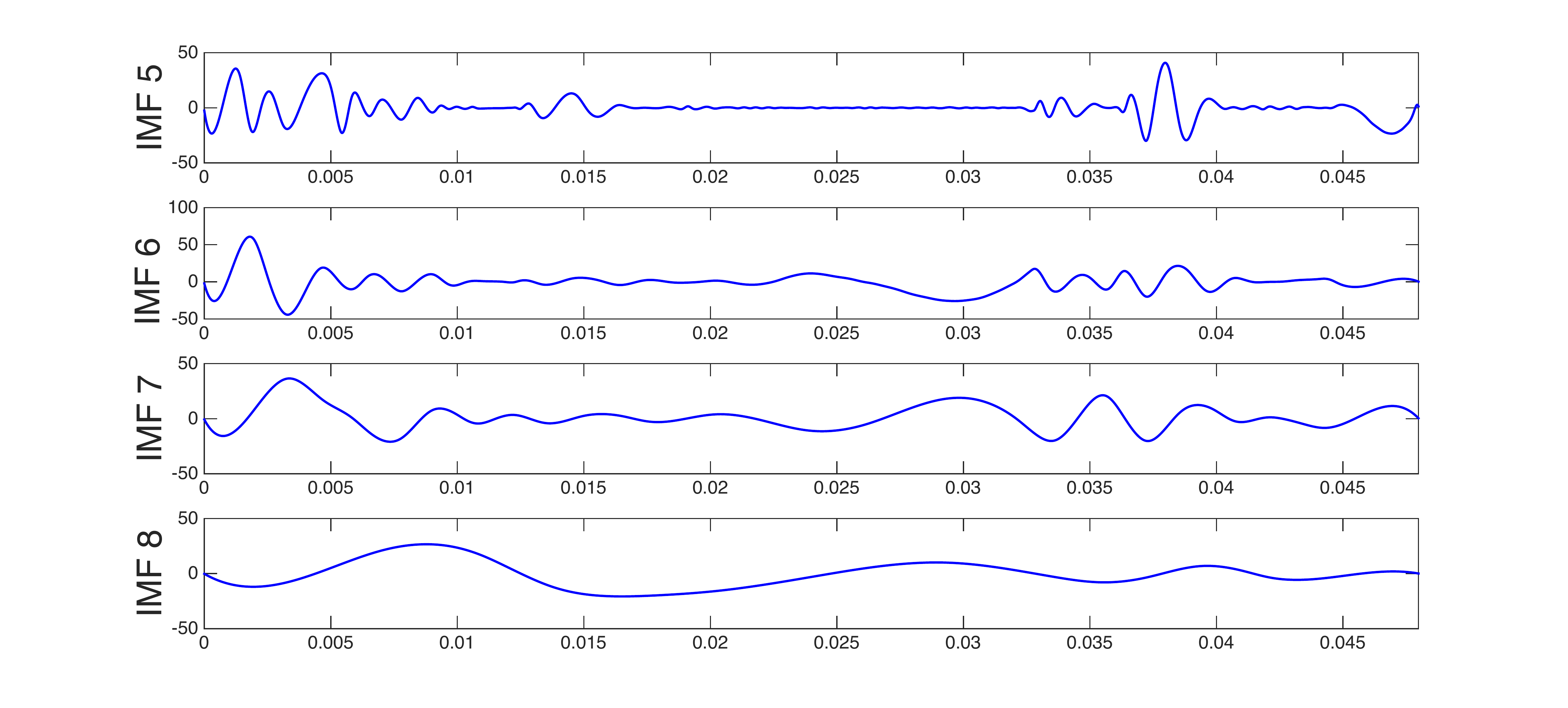}
\includegraphics[width = 0.45 \textwidth]{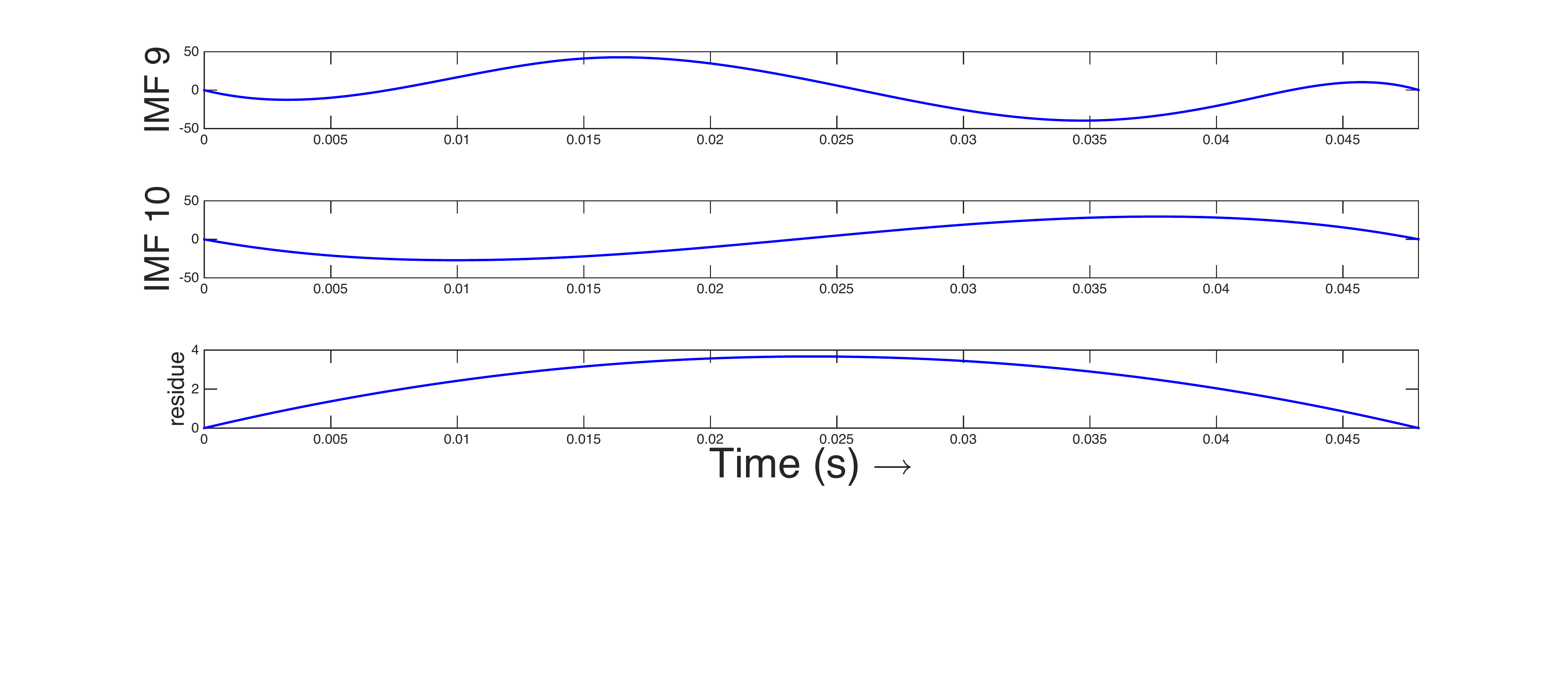}
\caption{\label{signal-imf} IMF components for the signal obtained from decomposition}
\end{figure}

\begin{figure}
\centering
\includegraphics[width = 0.45 \textwidth]{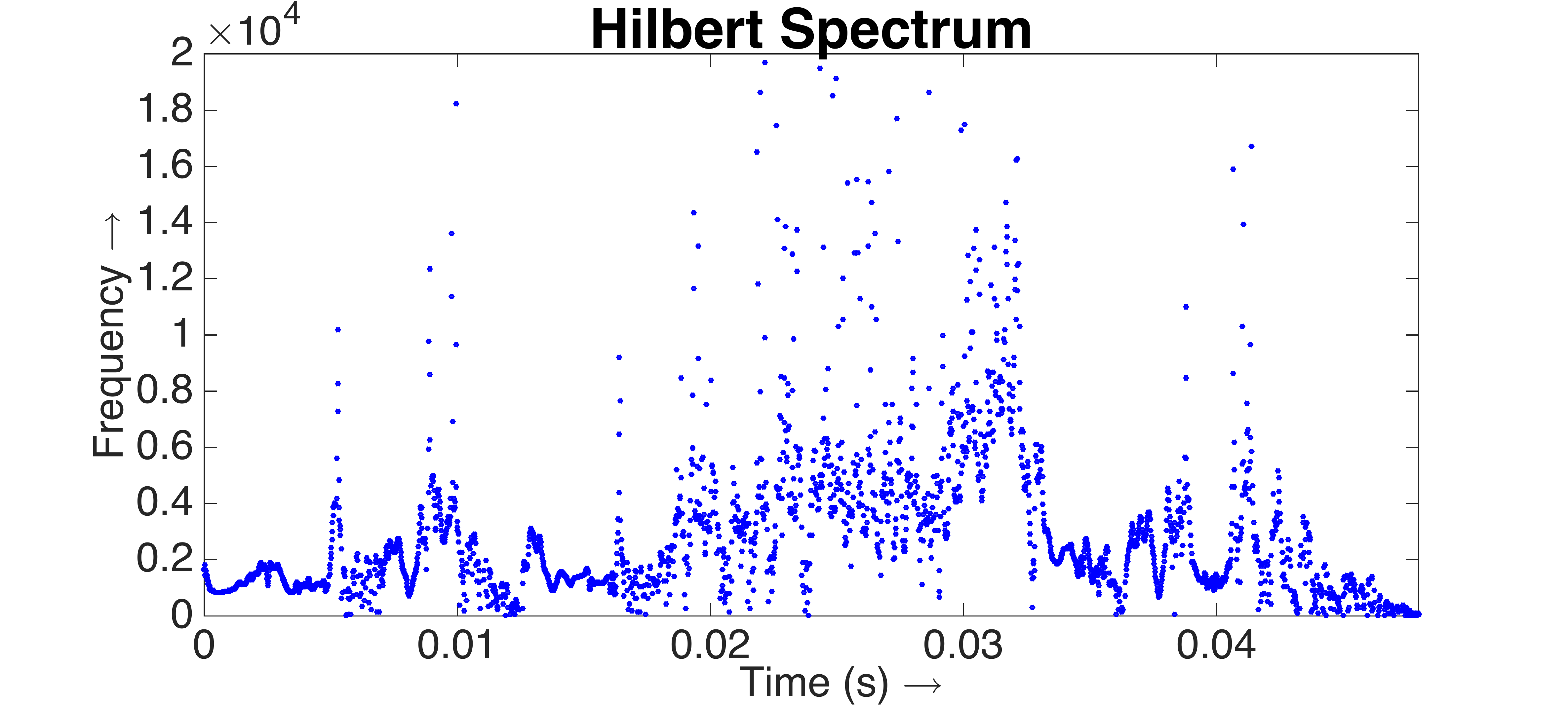}
\caption{Hilbert spectrum of the recorded signal.}
\end{figure}

The signals were first de-noised using the Butterworth filter to remove the unphysical frequency components. The de-noised signals were then analysed for their IMF components. This was followed by Hilbert spectrum analysis to achieve time-frequency localisation \cite{Huang08031998}.
FIG. \ref{signal-imf} depicts the IMF content of a typical signal, where the high frequency components are captured by the first few IMFs, while the higher ones represent the low frequency modulations. It is evident from FIG. \ref{signal_emd} and FIG. \ref{signal-imf} that the ruptures are well represented by the initial two IMFs, while the others capture the transient periodic as well as low frequency modulations, showing the effectiveness of the Hilbert-Huang transformation.
\par
The ability to single out such low frequency pulse like (LFPL) signals with transience and knowing the time-frequency distribution can aid in investigating the source mechanism and the extent of nonlinearity of the system. We now proceed to study the possible occurrence of SOC like behaviour in this system.

\section{Self Organised Criticality}
SOC has manifested in simple dynamical systems, like the Bak and Sneppen (BS) model \cite{PhysRevLett.71.4083}. Ito successfully used this model to explain the SOC in earthquakes \cite{PhysRevE.52.3232}.
\par
Ruptures having a magnitude greater than the power input to the system are treated as avalanches. It is to be noted that the rise in power originates from the impact of the balls on the piezoelectric layer. After impact, the balls lose contact with the layer and bounce back. Another avalanche follows when a ball strikes again. This is analogous to the fault model \cite{PhysRevE.52.3232}, where a rupture starts from the site with minimum barrier strength. After the site breaks, new random numbers are assigned to that site and its two adjoining neighbours. Rupture propagates as long as the new barriers are weaker than the threshold value of the rupture. The threshold for our setup is the power of the input signal used for the system.
\begin{figure}
\centering
\includegraphics[width = 0.45 \textwidth]{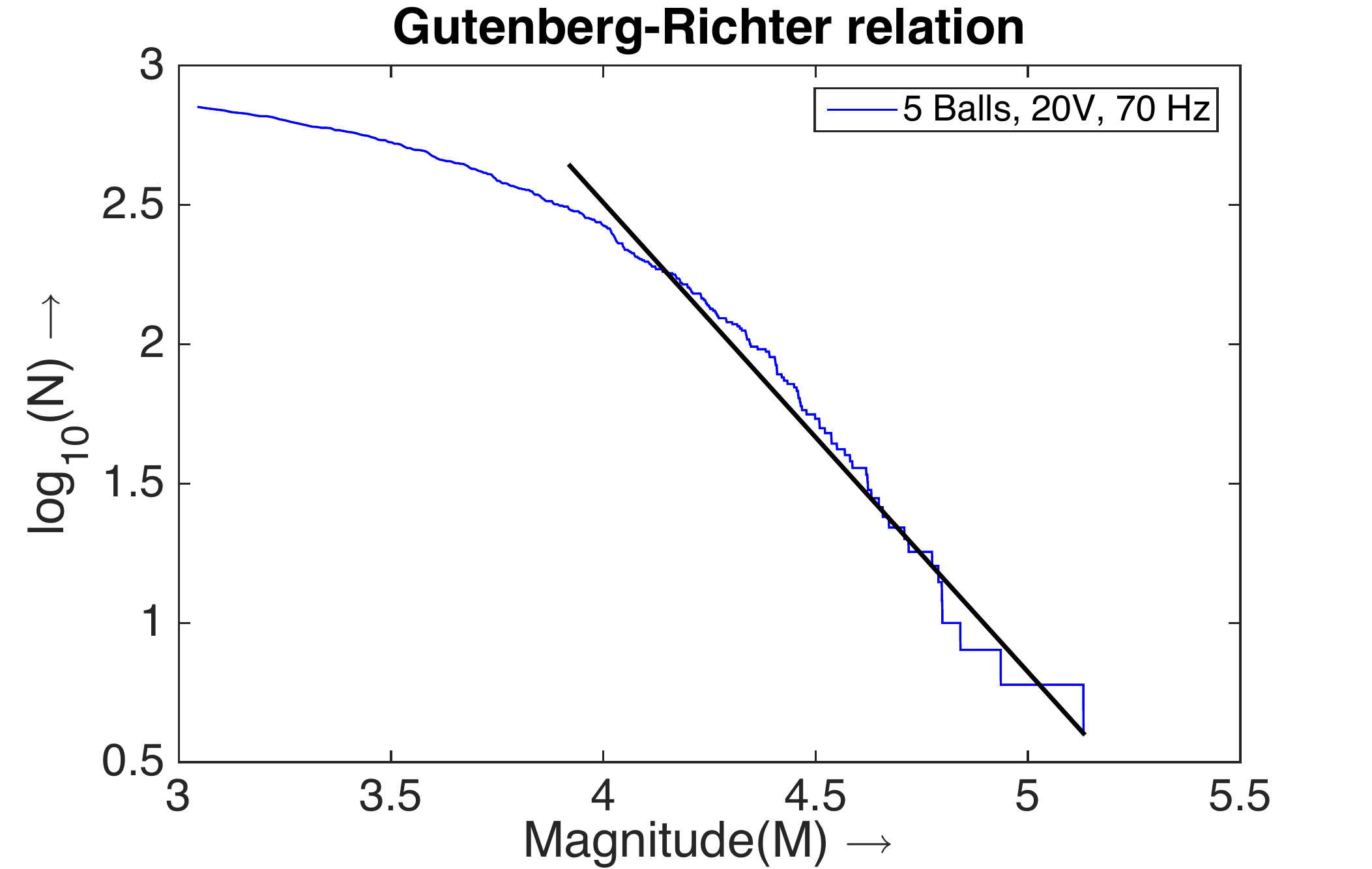}
\caption{\label{gr_reln}Emergence of Gutenberg-Richter type relation for a typical configuration of the system with, $a$ and $b$  as $-1.68$ and $9.25$ respectively, indicating larger number of minor ruptures as compared to major ones.}
\end{figure}
\begin{figure}
\centering
\includegraphics[width = 0.45\textwidth]{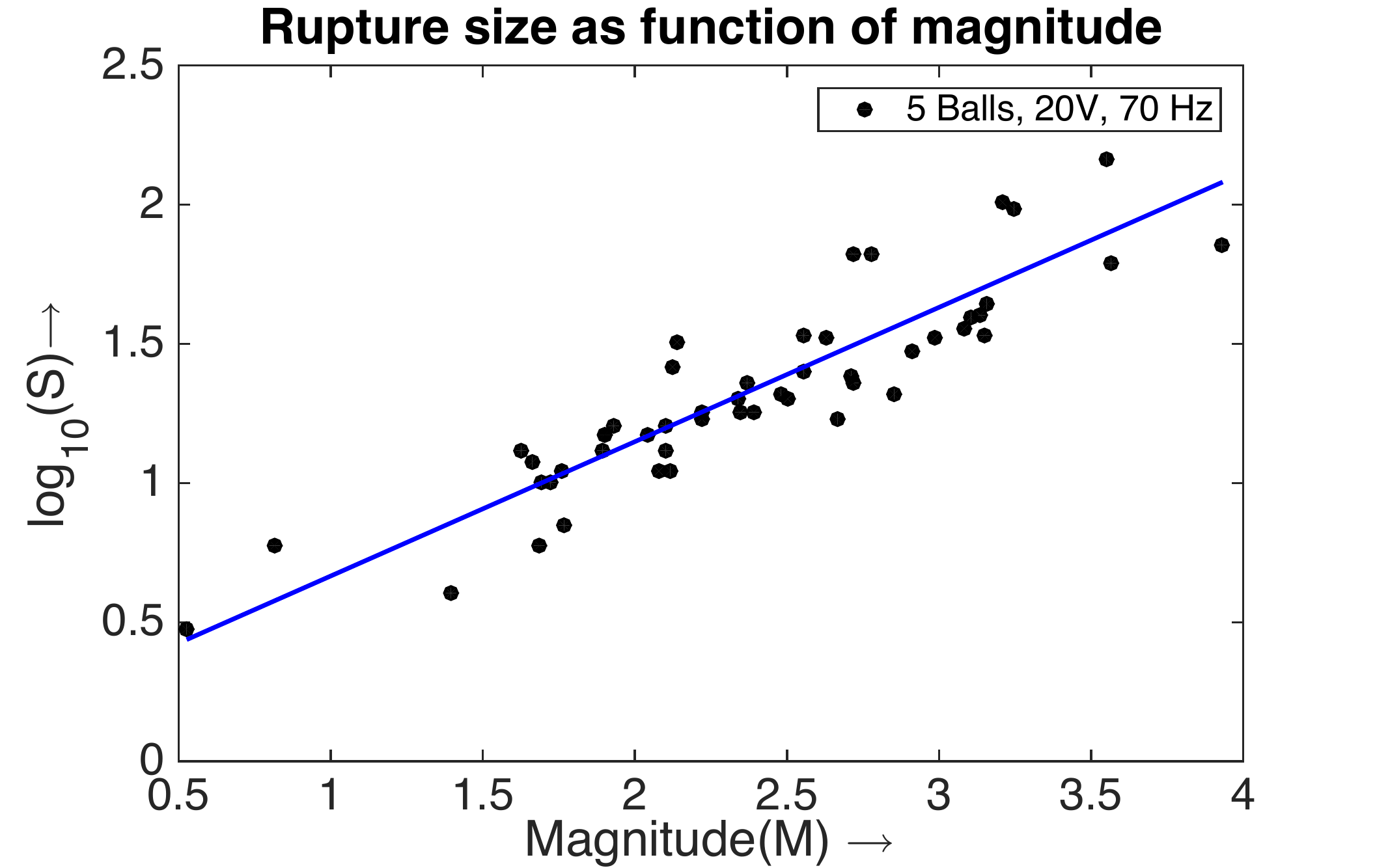}
\caption{\label{sm_reln}Logarithm of rupture size as a function of magnitude for a typical configuration of the system with $c$ and $d$ as $0.48$  and $0.18$ showing linear behaviour.}
\caption{Gutenberg-Richter relation and logarithm of rupture size as a function of magnitude.}
\end{figure}
\par
Keeping in mind the similarity of the time series with the ones connected with earthquakes, we begin with possible observation of Gutenberg-Richter \cite{gr_law} type relation in the present case. For this purpose the frequency of occurrence of the ruptures in the time series was computed. FIG. \ref{gr_reln} depicts a power law distribution in the frequency of occurrence of ruptures as a function of their power, proportional to the square of the amplitude. Evidently, the power quantifies the magnitude of the rupture.
The observed behaviour is similar to the Gutenberg-Richter \cite{gr_law} relation:
\begin{equation}
log_{10}N(>M) = a - bM.
\end{equation}
In the present case, $M$ is the magnitude of the rupture and $N(>M)$ is the number of ruptures with the magnitude greater than $M$. We also observed that the value of $b$ is lesser for events with smaller magnitude; this represents the well known `roll-off', a prominent feature of the Gutenberg-Richter plots.
It has also been established that there is a linear relation between the logarithm of the size $S$ of the ruptured zone and its magnitude  \cite{sm_reln}:
\begin{equation}
log_{10}S = cM + d.
\end{equation}
and thus
\begin{equation}
N \sim S^{-\frac{b}{c}}.
\end{equation}
\par
To observe the relation between size and magnitude of a rupture, we have considered the time duration for which the rupture lasts as a measure of its size. Moreover the time duration of the entire signal has been scaled to the number of bins for which it was recorded. The size and the magnitude of the rupture follows a linear relation as seen clearly in FIG. \ref{sm_reln}.

It is worth pointing out that, the presence of power-law behaviour is not adequate to claim that the system exhibits ``punctuated equilibrium". However, the presence of power-law correlations is considered to be a satisfactory proof \cite{PhysRevE.51.1059}. For this purpose, we define the temporal correlation function $P_t(t)$ as the probability that the time difference between the initiation of two ruptures is $t$ and analyse its behaviour as a function of time difference between occurrence of two ruptures. Our results clearly show that the correlation function indeed obeys power-law behaviour:
$$ P_t(t) \sim t^{-m}$$,
$m$ being a positive constant. The exponent for $P_t(t)$ obtained was less than $1$, satisfying the requirement of criticality as shown in \cite{PhysRevE.51.1059}.

\begin{figure}
\centering
\includegraphics[width = 0.45 \textwidth]{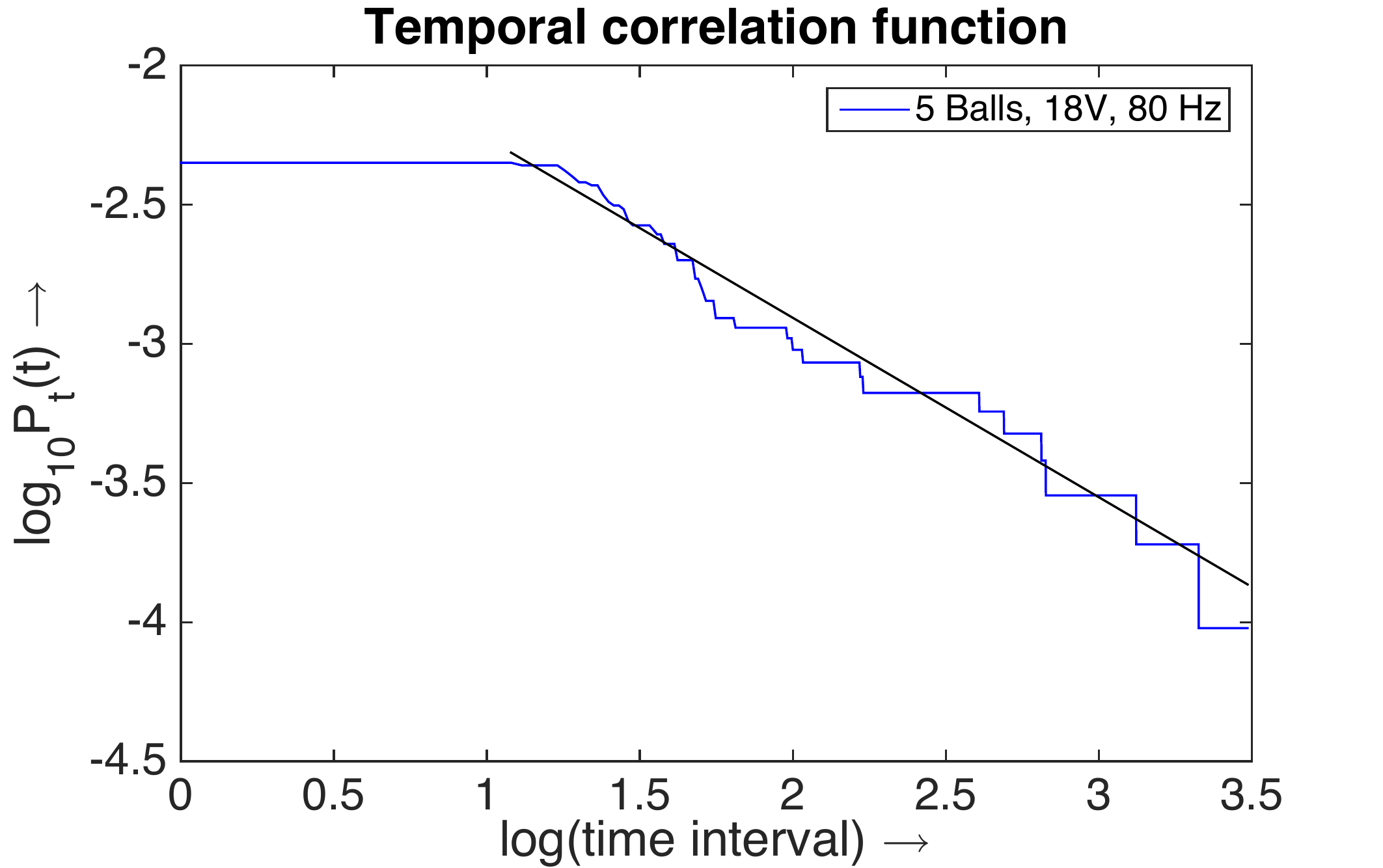}
\caption{\label{corr_t}Temporal correlation function for a typical configuration of the system, $P_t(t) \sim t^{-0.65}$, showing power law behaviour, with exponent less than one.}
\end{figure}

We now turn our attention to possible occurrence of Devil's staircase as has been observed in other models showing SOC and PE like behaviour. For the purpose of clarity and comparison with the present case we briefly outline Bak's model \cite{PhysRevE.53.414}.
\section{The Devil's Staircase}
Bak in his model of evolution had represented the accumulated mutation for certain species over time. Intercalated periods of stasis and avalanches of changes were observed for this species. The period after the motion for which the value of accumulated mutation remained constant showed stasis and the vertical jumps representing mutations indicated the avalanches of change. Such curves with many small jumps and a few big jumps are called the `Devil's staircases'. Bak established that this curve is a good explanation for the data of mutation in the physical size of some species.
\par
Devil's staircase can also manifest in simpler systems. Bak and Bruinsma \cite{PhysRevLett.49.249} have proven that the one-dimensional Ising model with long-range antiferromagnetic interactions exhibits a complete Devil's staircase. Boettcher and Pacuzki \cite{PhysRevLett.76.348}, introduced yet another SOC model of punctuated equilibrium with many internal degrees of freedom ($N$) per site. They found exact solutions for $N \rightarrow \infty$ of cascade equations describing avalanche dynamics in the steady state and successfully illustrated that punctuated equilibrium is described by a Devil's staircase.
\par
\begin{figure}
\centering
\includegraphics[width = 0.45 \textwidth]{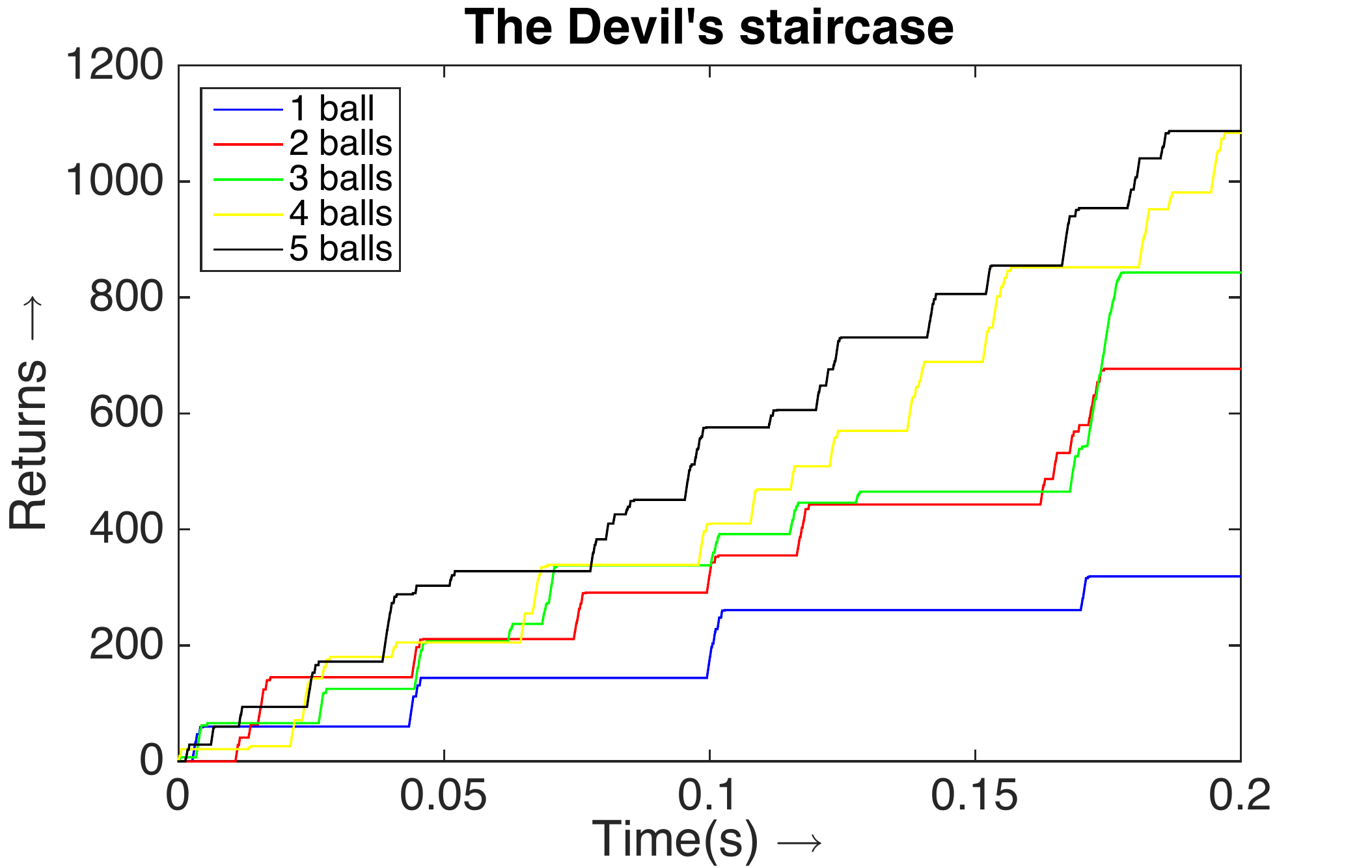}
\caption{\label{staircase}Returns as a function of time showing emergence of Devil's staircase for different number of balls where $V = 10~V$, $\omega = 70~Hz$}
\end{figure}

In our analysis, we have assumed that the state in which the power is lesser than the input power is a state of stability or stasis. It has been already suggested that criticality emerges via self-organising processes which provide stability \cite{PhysRevLett.115.208103,1478-3975-12-1-016001}. Returns to stasis state were plotted as a function of time. As is illustrated by FIG. \ref{staircase}, it gave rise to Devil's staircase which became prominent as the number of balls increased.
\par
It is evident that for a single bouncing ball, there are long periods of stasis with small jumps, spread wide apart in time. It is clearly seen for the case of four or five number of balls that the returns plotted as a function of time indeed show behaviour akin to Devil's staircase. The jumps become more frequent and the period of stasis lessens.
\par
A quantitative investigation will require modelling of bouncing balls through appropriate dynamical equations. For the single ball the same has been accomplished  \cite{PhysRevLett.65.393, PhysRevE.48.3988}. This needs to be extended for multiple balls.
\section{Conclusion}
In conclusion, we have demonstrated that the system of bouncing balls indeed exhibits features similar to self-organised-criticality and punctuated equilibrium behaviour. The Hilbert-Huang transformation effectively isolates transience and low frequency modulations in this non-stationary time series, as well as achieving time-frequency localisation, which are essential for unravelling these features. The Devil's staircase emerged prominently in case of larger number of balls indicating that an interacting system with multiple degrees of freedom leads to this behaviour. This motivates us to study the underlying dynamical system by generalising the case of a single bouncing ball to multiple scenario. This nonlinear coupled system needs numerical investigation. We hope to report on this in the near future.
\section*{Acknowledgments}
K.G. acknowledges Abhinna K. Behera and Pankaj Kumar Shaw, for insightful discussions during the course of this work. K.G. is also thankful to IISER-Kolkata and SINP for their hospitality.

\bibliography{references}

\end{document}